# DEPLOYMENT AND SIMULATION OF THE ASTROD-GW FORMATION

AN-MING WU

*National Space Organization (NSPO), Hsinchu, Taiwan, 300, ROC*
*amwu@nspo.narl.org.tw*

WEI-TOU NI

*Center for Gravitation and Cosmology (CGC)*
*Department of Physics, National Tsing Hua University, Hsinchu, Taiwan, 300, ROC*
*weitou@gmail.com*

Constellation or formation flying is a common concept in space Gravitational Wave (GW) mission proposals for the required interferometry implementation. The spacecraft of most of these mission proposals go to deep space and many have Earthlike orbits around the Sun. ASTROD-GW, Big Bang Observer and DECIGO have spacecraft distributed in Earthlike orbits in formation. The deployment of orbit formation is an important issue for these missions. ASTROD-GW (Astrodynamical Space Test of Relativity using Optical Devices optimized for Gravitation Wave detection) is to focus on the goal of detection of GWs. The mission orbits of the 3 spacecraft forming a nearly equilateral triangular array are chosen to be near the Sun-Earth Lagrange points L3, L4 and L5. The 3 spacecraft range interferometrically with one another with arm length about 260 million kilometers with the scientific goals including detection of GWs from Massive Black Holes (MBH), and Extreme-Mass-Ratio Black Hole Inspirals (EMRI), and using these observations to find the evolution of the equation of state of dark energy and to explore the co-evolution of massive black holes with galaxies. In this paper, we review the formation flying for fundamental physics missions, design the preliminary transfer orbits of the ASTROD-GW spacecraft from the separations of the launch vehicles to the mission orbits, and simulate the arm lengths of the triangular formation. From our study, the optimal delta-Vs and propellant ratios of the transfer orbits could be within about 2.5 km/s and 0.55, respectively. From the simulation of the formation for 10 years, the arm lengths of the formation vary in the range $1.73210 \pm 0.00015$ AU with the arm length differences varying in the range $\pm 0.00025$ AU for formation with 1° inclination to the ecliptic plane. This meets the measurement requirements. Further studies on the optimizations of deployment and orbit configurations for a period of 20 years and with inclinations between 1° to 3° are currently ongoing.

## 1. Introduction

The definition of formation flying given by NASA Goddard Space Flight Center is: The tracking or maintenance of a desired relative separation, orientation, or position between or among spacecraft.[1] Examples in geodesy are GRACE[2] mission and GOCE[3] mission. A dense formation for high resolution interferometry and making maps of Earth is TechSat-21.[4] On the other hand, a constellation consists of a number of spacecraft with coordinated coverage and operations, like GPS[5], Iridium[6], DMC[7], and FORMOSAT-3.[8] As an example, FORMOSAT-3 is a constellation of 6 satellites deployed in six orbital planes with equally separated right ascensions to gather the global meteorological data.[8]

In a way, SLR (Satellite Laser Ranging) and LLR (Lunar Laser Ranging) are simplest kind of constellations with satellite(s) and laser station(s) on Earth forming the constellation. The ranges coming out of the missions serve as science data for applications to geophysics, reference frames, selenophysics, fundamental physics etc. ASTROD I with laser ranging to a deep space drag-free spacecraft to map the solar system and to explore fundamental physical laws belongs to this category also.[9]

Laser-interferometric gravitational wave (GW) detectors use constellation formation in deep space to compare dynamic arm changes in order to detect and measure GWs. These space missions include LISA,[10] NGO,[11] DECIGO,[12,13] BBO,[14] ASTROD-GW[15,16] and Super-ASTROD.[17]

Laser Interferometer Space Antenna (LISA) is a proposed space mission concept designed to detect and accurately measure gravitational waves from astronomical sources. LISA Pathfinder is designed for a proof-of-concept mission, and is due for launch in 2015.[18] LISA has three spacecraft arranged in a nearly equilateral triangle formation with 5 million kilometer arms, inclined by 60˚ with respect to the ecliptic and flying along an Earth-like heliocentric orbit trailing Earth by 20˚. It would be sensitive to the GWs in the frequency band of 0.03 mHz ~ 100 mHz.

New Gravitational wave Observatory (NGO/eLISA [evolved LISA]) mission is derived from the previous LISA proposal, and will survey in the low-frequency GW band of 0.1 mHz ~ 1 Hz. NGO/eLISA constellation comprises three spacecraft operating in a V formation, with the mother spacecraft having two free-falling test masses that define the vertex of the two interferometer arms while the other two daughter spacecraft the end-points of the interferometer arms. The three spacecraft will orbit the Sun and have a near-equilateral triangular formation with an arm length of 1 million kilometers, inclined by 60˚ with respect to the ecliptic and flying along an Earth-like heliocentric orbit trailing Earth and gradually drifting away from 10˚ to 20˚.

DECIGO (DECi-hertz Interferometer Gravitational wave Observatory) is a space GW antenna with the purpose to observe gravitational waves at the frequency band of 0.1 Hz ~ 10 Hz.[12] DECIGO will consists of four clusters. Each cluster will have 3 drag-free spacecraft forming a triangle similar to that of LISA configuration in solar orbit but with arm lengths maintained at 1000 km. Each arm forms a Fabry-Perot cavity instead of being transponder-type. 3 clusters are distributed 120˚ apart in 1 AU orbits with the remaining triangular cluster on top of one of the 3 clusters, in a hexagram formation. DECIGO pathfinder (DPF)[13,19] is the first milestone mission to test the key technologies of DECIGO with one spacecraft. It will make observations at 0.1 Hz ~ 1 Hz, using a small satellite with weight about 350 kg on a Sun-synchronous orbit of 500 km.

Big Bang Observer (BBO) is a proposed successor to LISA. The primary scientific goal will be the observation of GWs from the time shortly after the Big Bang, but it will also be able to detect younger sources of gravitational radiation, like binary inspirals. BBO will bridge the frequency detection gap between adLIGO[20,21] and NGO/eLISA. BBO is a collection of four LISA-like formations with arm length 50,000 km, each composed of three spacecraft flown in a triangular pattern. Two of the triangles will be on top of each other, in a hexagram formation. The other two triangles will be located at distant places along Earth's orbit.

ASTROD-GW (ASTROD [Astrodynamical Space Test of Relativity using Optical Devices] optimized for Gravitation Wave detection) is an optimization of ASTROD[22] to focus on the goal of detection of GWs. The detection sensitivity is shifted toward larger wavelength compared to that of LISA and NGO/eLISA. The mission orbits of the 3 spacecraft forming a nearly equilateral triangular array are chosen to be near the Sun-Earth Lagrange points L3, L4, and L5 (Fig. 1). The 3 spacecraft range interferometrically with one another with arm length about 260 million kilometers.[15,16] With 52 and 260 times longer in arm length compared to that of LISA and NGO/eLISA respectively, the strain detection sensitivity is 52 and 260 times better at large wavelength. The scientific aim is focused for GW detection at low frequency. The science goals include detection of GWs from Massive Black Holes (MBH), and Extreme-Mass-Ratio Black Hole Inspirals (EMRI), and using these observations to find the evolution of the equation of state of dark energy and to explore the co-evolution of massive black holes with galaxies. It will be sensitive to the gravitational waves in the frequency band of 100 nHz ~ 1 mHz.

All the above mission proposals use drag-free navigations. Particle detectors are proposed to monitor the charging process of the freely floating proof-masses. These particle detectors will naturally provide SEP (Solar Energetic Particle) observations at different helio-longitude and distances from Earth for space weather applications.[23]

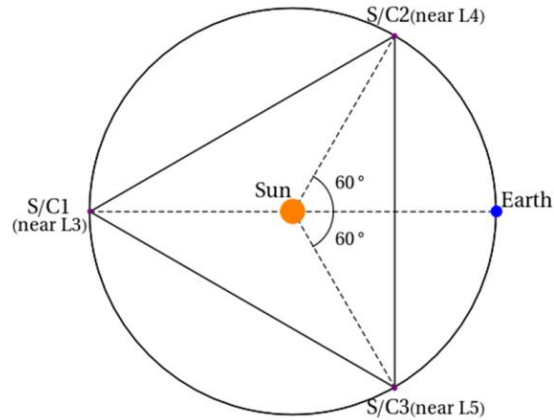

Fig. 1. Schematic of ASTROD-GW constellation formation.

Besides space interferometric constellations for GW detection, there are constellations including natural celestial bodies which are promising for GW detection. These constellations are Pulsar Timing Arrays (PTAs).[24-28] Pulsars emit radio pulses regularly and are ultra-stable clocks through precise timing of these radio pulses.[24,25] When very low frequency GWs (300 pHz - 100 nHz)[15,16,29,30] pass by the line of sight of pulsars, they encode periodic signals on the arrival times of pulses and therefore an array of pulsars emitting pulses with ultra-stable periods together with a radio telescope or an array of radio telescopes on Earth or in space form a constellation that can serve as a GW detector.[24,25]

In section 2, we work out the deployment of the ASTROD-GW formation. Since the total weights of the ASTROD-GW spacecraft are crucial for cost estimation of the mission, we must design and optimize the transfer orbits and evaluate the propellant mass ratios in the beginning phase of study to find out a safe requirement. The propellant mass ratio means the propellant mass shall be carried in portions of the spacecraft mass, and it relates to the delta-V of the impulse of propulsion for transfer orbit maneuvers. This will give an estimation range of the propellant mass ratios for DECIGO and BBO also. In section 3, we optimize the ASTROD-GW formation in the science phase, i.e., minimize the arm length variations and Doppler (relative line of sight) velocities between pairs of spacecraft. In the section 4, we discuss method of Venus swing-by, estimate the possible reductions of the delta-Vs and deployment time, and present an outlook.

**2. Deployment of the ASTROD-GW Formation**

To deploy the formation, we design the transfer orbits of the spacecraft from the separations of the launch vehicles to the mission orbits.

Each spacecraft is propelled by a high efficient propulsion module for large delta-V maneuvers and for delivery to the destination. This module is to be separated when the destination is achieved.

Lagrange points are the equilibrium points of the restricted three-body problem. *For two celestial bodies in circular orbits, the Lagrange points are stationary with respect to the two celestial bodies*. There are five Lagrange points, denote by L1~L5. The three points L1, L2, L3 are located on the line connected the two celestial bodies with L1 and L2 near the small body and L3 far away from the two celestial bodies. L4 and L5 are located out of the line connected the two celestial bodies.

*For two celestial bodies in elliptical orbits like the Sun and the Earth, the Lagrange points are not stationary with respect to the two celestial bodies.* However, for the Sun-Earth system, the eccentricity $e$ is 0.0167, and the relative position variation of Lagrange points are of this order too. The circular orbit of a spacecraft (S/C) around the Sun near L4 or L5 points will remain circular to $O(e^2)$. The spacecraft orbits near L3 point of the Sun-Earth system are unstable. However, the instability time scale is over 50 years. For a mission of 10-20 years, the orbits are virtual stable or quasi-stable. Therefore, we have chosen the nominal mission orbits of the ASTROD-GW spacecraft to be circular orbit in the heliocentric system near L3, L4, and L5 of the Sun-Earth system to form a nearly regular triangle with arm lengths about 1.732 AU.[15,16,30] The orbits could have small inclination angles of order of $O(e)$ too (see Sec. 3) and still achieve equilateral to $O(e^2)$. Planetary perturbations are considered in Sec. 3.

In Sec. 2.1 and Sec. 2.2, we use circular orbits of Sun and Earth as an approximation. In the Sun-Earth rotating frame in this approximation, position of the Sun and the Earth are fixed and optimization is simplified.

**2.1 Design of Transfer Orbits**

For the three spacecraft travelling from Earth to separate Lagrange points, we begin with the construction of two-impulse Hohmann transfer orbits with Sun attraction only for guidance. The starting point of the Hohmann transfer orbit is at the point on GEO (geosynchronous orbit) facing the Sun for S/C 1 and 2, and opposite to the Sun for S/C 3. We also first specify travel time for spacecraft 1 to L3 in 1+1/2 years or in 2+1/2 years, spacecraft 2 to L4 in 1+5/6 years, and spacecraft 3 to L5 in 2+1/6 years. From the travel time, we can determine the elliptic transfer orbits in the Sun-Earth barycentric frame as shown in Fig. 2. The orbits of Sun, Earth, and spacecraft are plotted in the figures. The initial positions are denoted by small circles, and the final positions by large circles. The two impulses are also shown with red arrows. The first impulse is the small one to transfer to near Lagrange points, and the second is to stay at the Lagrange points. The launch position is at right of the orbit. After appropriate transfer time, the Earth location is shown and the spacecraft returns to the original position.

With this guidance as reference, the transfer orbits under the attraction of Sun and Earth can be realized through the proper orbit design and control. The initial position and velocity are selected as those of geosynchronous orbits (GEO), so that each spacecraft is required to escape from the attraction of Earth and to enter into an orbit around Sun.

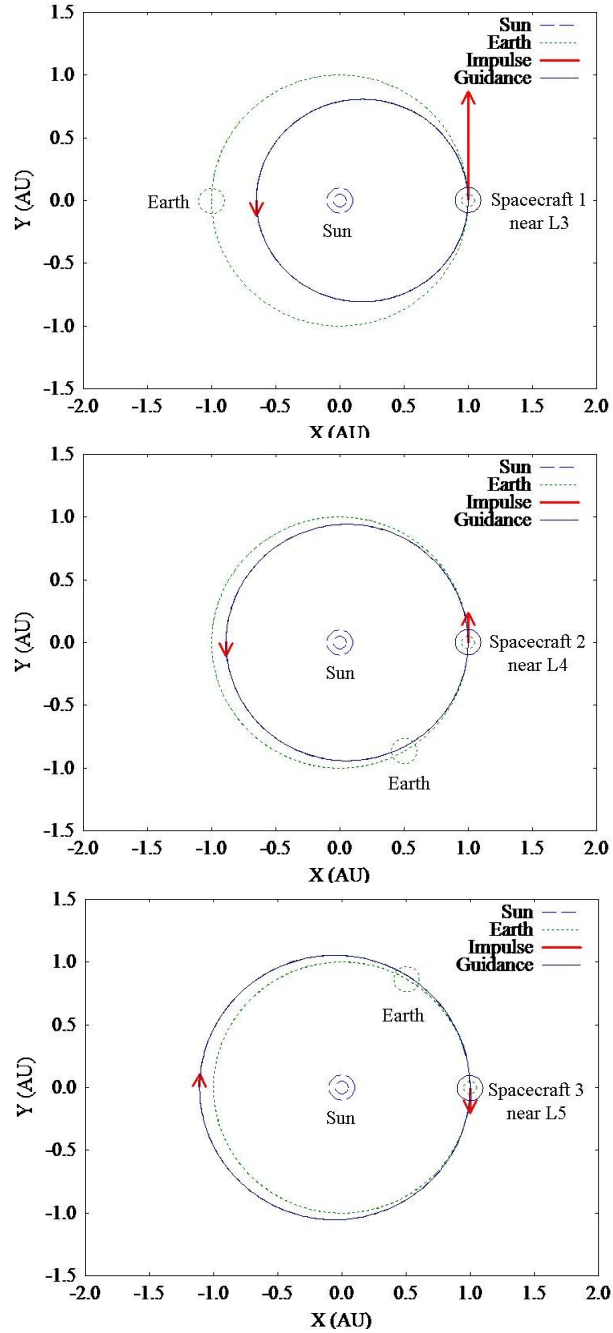

Fig. 2. ASTROD-GW S/C transfer orbits for S/C 1 from inner GEO to near L3 (top), S/C 2 from inner GEO to near L4 (middle), and S/C 3 from outer GEO to near L5 (lower).

**2.2 Calculation of Delta-Vs**

Consider the barycentric frame of the Sun-Earth system. The equation of motion of the spacecraft is as follows:

$$\frac{d^2\vec{r}}{dt^2} = \frac{Gm_s}{|\vec{r}_s - \vec{r}|^3}(\vec{r}_s - \vec{r}) + \frac{Gm_e}{|\vec{r}_e - \vec{r}|^3}(\vec{r}_e - \vec{r}), \qquad (1)$$

where $G$ is the gravitational constant, $\vec{r}$ the position of spacecraft, $\vec{r}_s$ the position of Sun, and $\vec{r}_e$ the position of Earth.

With two-point boundary values given, this equation can be solved accurately by the 4th-order compact difference method with tri-diagonal algorithm (TDMA) iteration.[31,32] The force terms are further linearized to overcome the difficulties for the spacecraft to escape from the Earth. The finite difference equation is then written as follows:

$$[\frac{\delta_t^2 \vec{r}}{\Delta t^2(1+\delta_t^2/12)}]^{new} = [\frac{Gm_s}{|\vec{r}_s - \vec{r}|^3}(\vec{r}_s - \vec{r}) + \frac{Gm_e}{|\vec{r}_e - \vec{r}|^3}(\vec{r}_e - \vec{r})]^{old}$$
$$+ \nabla[\frac{Gm_s}{|\vec{r}_s - \vec{r}|^3}(\vec{r}_s - \vec{r}) + \frac{Gm_e}{|\vec{r}_e - \vec{r}|^3}(\vec{r}_e - \vec{r})]^{old}.(\vec{r}^{new} - \vec{r}^{old}), \qquad (2)$$

where $\delta_t^2 \vec{r}_i \equiv \vec{r}_{i+1} - 2\vec{r}_i + \vec{r}_{i-1}$, $\Delta t$ and $i$ are respectively the time step and the index for the uniform partition of the time domain, and $\vec{r}_i$ is the position at time $t_i$.

The time domain is taken for the first half revolution of the transfer orbit. After the position vector solved by the above equation, the initial velocity is calculated by the 4th-order difference formula. The delta-V to escape from the Earth is taken as the difference of the initial velocity and the spacecraft velocity on GEO orbit. The computed results are shown in Fig. 3.

The various cases of delta-Vs and the propellant ratios of the three spacecraft are summarized in Table 1. For calculation of the propellant ratios we assume the specific impulse is 320 sec as used for NGO spacecraft design.[11] Note that the propellant mass ratio for spacecraft 1 from inner GEO to L3 in 1.500 years is 0.822, which is greater than those of the other two spacecraft. If it takes 2.500 years to transfer, then the ratio decreases to 0.681. Therefore, the propellant mass ratios of three spacecraft can all be smaller than 0.7. From this number we can estimate the total mass of each spacecraft. If the dry mass of the spacecraft is 500 kg, which counts the payload and all subsystems, then the total mass of each spacecraft will be 2174 kg, assuming the propulsion module takes 10% mass of the propellant.

In the mission study of ASTROD I,[33-35] the ASTROD I spacecraft is given an appropriate delta-V before the last stage of launcher separation in the LEO (Low Earth Orbit) and is injected directly to the solar orbit going geodetic to Venus swing-by. This means that we should use the same strategy to launch the ASTROD-GW spacecraft directly into the solar transfer orbits near the designated Hohmann orbits. This way, only the solar transfer part of delta V is needed for each spacecraft to reach the destination. Most of this delta V needed occurs near the destination to boost the spacecraft to stay near the destined Lagrange point. In the last 2 columns of Table 1, we list the values of solar transfer delta-V and propellant mass ratio. In this case, we obtain the propellant mass ratios around 0.470 (2.5 year transfer to near L3), 0.250 and 0.216 for spacecraft 1, 2 and 3, and the total mass 1035 kg, 690 kg, and 656 kg corresponding to the dry spacecraft (not including propeller and propellant) mass 500 kg.

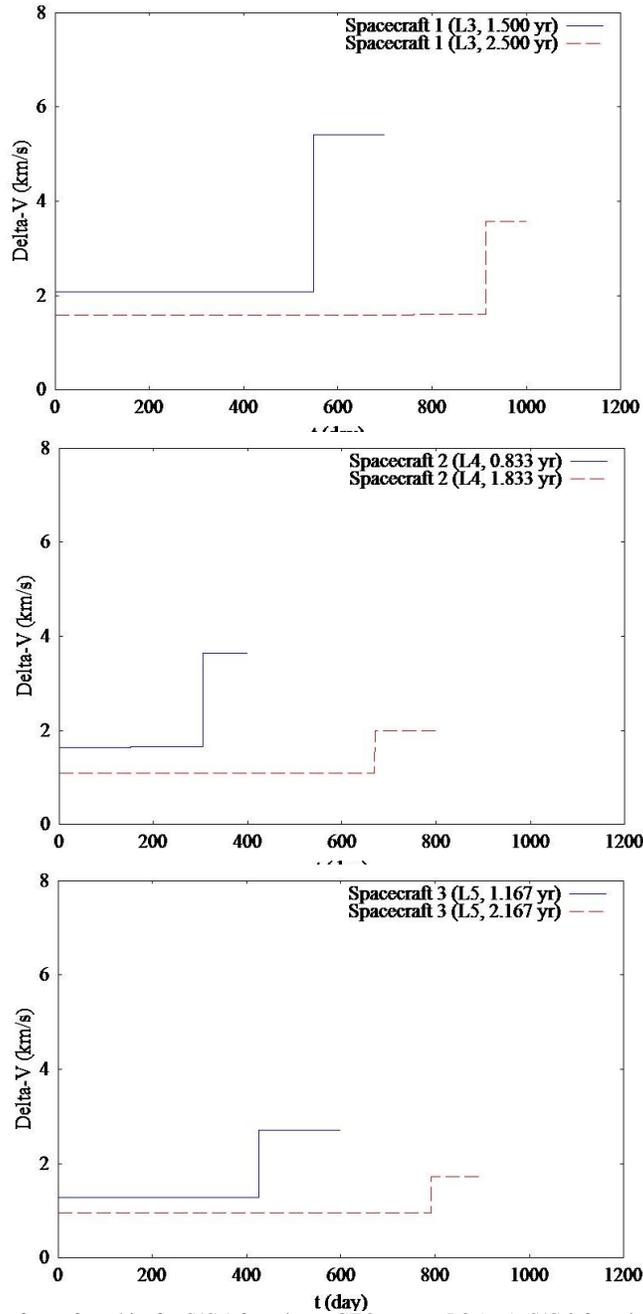

Fig. 3. Delta-Vs of transfer orbits for S/C 1 from inner GEO to near L3 (top), S/C 2 from inner GEO to near L4 (middle), and S/C 3 from outer GEO to near L5 (lower)

Table 1. Delta-V and Propellant Mass Ratio

| S/C (Destination) | Transfer Orbit | Total Delta-V [km/s] | Total Propellant Mass Ratio (Isp=320s) | Delta-V to Enter the Solar Orbit from GEO [km/s] | Solar Transfer Delta-V from Beginning of Hohmann Orbit to Destination [km/s] | Solar Transfer Propellant Mass Ratio (Isp=320 s) |
|---|---|---|---|---|---|---|
| 1 (near L3) | Inner Hohmann, 2 Revolutions in 1.500 Years | 5.410 | 0.822 | 2.075 | 3.335 | 0.655 |
| 1 (near L3) | Inner Hohmann, 3 Revolutions in 2.500 Years | 3.582 | 0.681 | 1.591 | 1.991 | 0.470 |
| 2 (near L4) | Inner Hohmann, 1 Revolution in 0.833 Years | 3.635 | 0.686 | 1.643 | 1.992 | 0.470 |
| 2 (near L4) | Inner Hohmann, 2 Revolutions in 1.833 Years | 1.992 | 0.470 | 1.089 | 0.903 | 0.250 |
| 3 (near L5) | Outer Hohmann, 1 Revolution in 1.167 Years | 2.706 | 0.578 | 1.284 | 1.422 | 0.365 |
| 3 (near L5) | Outer Hohmann, 2 Revolutions in 2.167 Years | 1.718 | 0.422 | 0.953 | 0.765 | 0.216 |

## 3. Simulation of the Arm Lengths

The spacecraft payload includes a drag-free system with micronewton thrusters to maintain the geodetic motion of the spacecraft in the science mode for GW detection after arriving destination.

To simulate the arm lengths and their differences in the formation of the three spacecraft, we compute the mission orbit evolutions in the solar system. The equation of motion is the same as Equation (1), but the force term now consists of the Newtonian or post-Newtonian attraction of all the planets and the Moon in the solar system. The initial conditions of solar system are obtained from JPL DE405,[36] and the equations are integrated by 6th-order Runge-Kutta method for 10 years.

The basic configuration uses inclined circular orbit in the heliocentric ecliptic coordinate system. The orbit equation for an inclined circular orbit is

$$\begin{bmatrix} x' \\ y' \\ z' \end{bmatrix} = \begin{bmatrix} a[1 - \sin^2\Phi_0 (1-\cos\lambda)] \cos\varphi + a \sin\Phi_0 \cos\Phi_0 (1-\cos\lambda) \sin\varphi \\ a \cos\Phi_0 \sin\Phi_0 (1-\cos\lambda) \cos\varphi + a[1 - \cos^2\Phi_0 (1-\cos\lambda)] \sin\varphi \\ -a \sin\Phi_0 \sin\lambda \cos\varphi + a \cos\Phi_0 \sin\lambda \sin\varphi \end{bmatrix}, \qquad (3)$$

where $\lambda$ is the inclination angle, $\Phi_0$ is the right ascension of ascending node (RAAN), $\varphi = \omega t + \varphi_0$ with $\varphi - \Phi_0$ the true anomaly, and $a$ the semi-major axis corresponding to $\omega$ the mean motion of 1 rev/sidereal year.[15,16] The time $t$ is chosen to be zero at initial moment of simulation.

For the three orbits with inclination $\lambda$ (in radian), we choose:

S/C I: $\Phi_0(I) = 270°$, $\varphi_0(I) = 0°$,
S/C II: $\Phi_0(II) = 30°$, $\varphi_0(II) = 240°$,

S/C III: $\Phi_0$(III) = 150° $\varphi_0$(III) = 120°. (4)

The arm lengths are calculated to be

$|\underline{V}_{\text{II-I}}| = 3^{1/2} a [(1 - \xi/2)^2 + \sin^2 \lambda \sin^2 (\omega t + 60°)]^{1/2}$,

$|\underline{V}_{\text{III-II}}| = 3^{1/2} a [(1 - \xi/2)^2 + \sin^2 \lambda \sin^2 (\omega t)]^{1/2}$,

$|\underline{V}_{\text{I-III}}| = 3^{1/2} a [(1 - \xi/2)^2 + \sin^2 \lambda \sin^2 (\omega t - 60°)]^{1/2}$, (5)

with $\xi \equiv 1 - \cos \lambda$. The fractional arm length variation is within (1/2) $\sin^2 \lambda$ which is about $10^{-4}$ for $\lambda$ about 1°.[15,16]

The orbit simulation for the inclination angle equal to zero case has been worked out in previous studies.[37-41] From the simulation of the formation for 10 years, the arm lengths of the formation vary in the range 1.73205 ± 0.00015 AU with the arm length differences varying in the range ± 0.00025 AU for formation with no inclination to the ecliptic plane for the science orbit in 2025 using CGC 2.5 ephemeris[37,38] incorporating Brumberg's[42] post-Newtonian equation of motion. From the simulation of the formation for 20 years, the arm lengths of the formation vary in the range 1.73205 ± 0.00016 AU with the arm length differences varying in the range ± 0.0003 AU for formation with no inclination to the ecliptic plane for the science orbit in 2028.[41]

Here we worked out the orbit simulation for the inclination angle equal to 1° case. The main reason is that for resolving the antipodal ambiguity in the direction of detected GWs, we need to consider the mission orbits with a small inclinations.[15,16]

As a consistent check, we also compute the inclination angle equal to zero case taking the same initial time (at noon, June 21, 2025 [JD2460848.0]) and the same initial conditions of positions and velocities as in Ref. [37, 38]. In spite of using Newtonian ephemeris, the computed results of the evolution of arm lengths, and arm length differences are the same as in Fig. 4 of Ref. [37, 38] (Note that the labeling S/C is different in Ref. [37, 38]). The time dependence of the arm length differences is of the order of $10^{-3}$ fractionally as our previous calculations. The computed results for the case of 1° inclination case are shown in Fig. 4. For this case, the initial positions of 3 S/C remain the same as the no inclination case, the initial velocities are rotated up 1° to obtain the inclination. Fig. 4 looks almost identical to Fig. 4 of Ref. [37, 38] (Note that the labeling S/C is different in Ref. [37, 38]). This is expected since with 1° inclination, the arm lengths do not change by more than $O(10^{-4})$ fractionally. And the planetary perturbations also do not change the arm lengths by more than this order too. Hence these results are consistent with our previous results for the first simulation without optimization.[37-38]

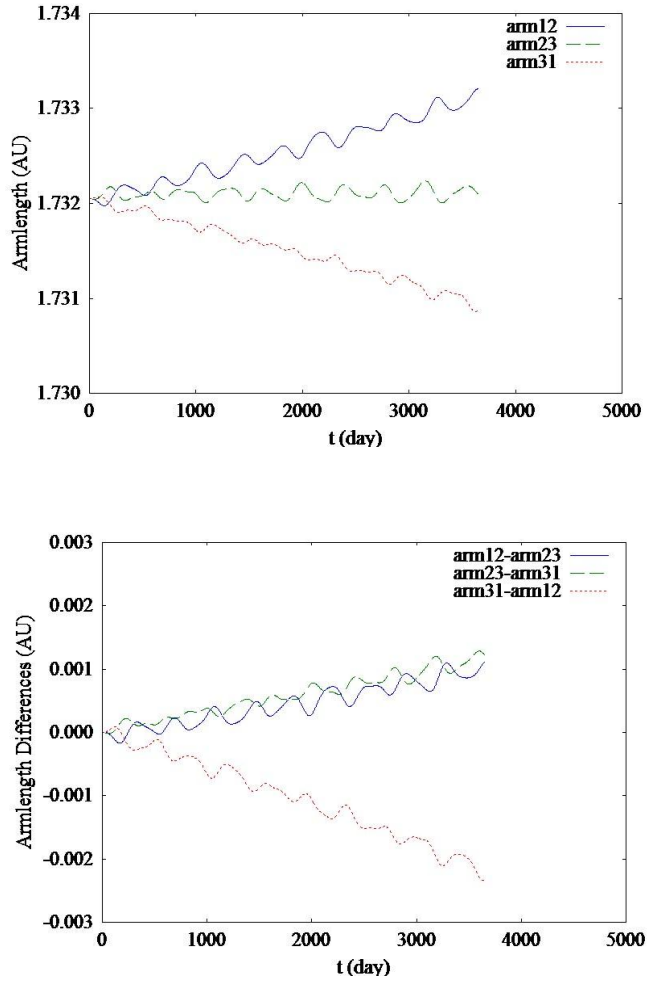

Fig. 4. Computed results of arm lengths (top) and arm length differences (bottom) for the case of 1° inclination from initial choice of initial conditions (at noon, June 21, 2025 [JD2460848.0]).

From Fig. 4, we see that S/C 1 is moving away from S/C 2 and moving toward S/C3 with the distance between S/C 2 and S/C 3 stay nearly constants. So we adjust the initial velocity components of S/C 1 in the ecliptic plane. After several trial adjustments, we find that the adjustment of the initial velocity of spacecraft 1 near L3 by the small increments of $\Delta u = 5 \times 10^{-7}$ AU/day and $\Delta v = -2 \times 10^{-7}$ AU/day will have the three arm lengths nearly equal and hence the mutual differences nearly zero. The computed results are shown in Fig. 5. The arm lengths of the formation vary in the range $1.73210 \pm 0.00015$ AU with the arm length differences varying in the range $\pm\,0.00025$ AU for this configuration with 1° inclination to the ecliptic plane. This meets the requirements of ASTROD-GW.

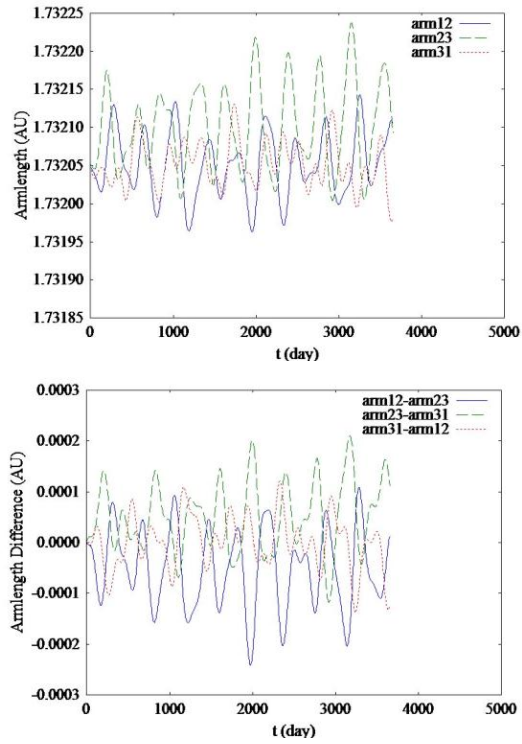

Fig. 5. Computed results of Lagrange points evolution (top), arm lengths (middle), and arm length differences (bottom) for the case of 1 deg inclination with adjustment of initial velocity.

The initial position x, y, z and initial velocity u, v, w (in this order) of the three S/C with $1°$ inclination for the initial choice and for the adjusted choice in barycentric rotating frame and barycentric inertial frame are shown in Table 2.

The differences of the arm lengths of the triangular formation for three S/C with $1°$ inclination have been simulated to meet the measurement accuracy requirements. In this first simulation of inclined orbits, the effects of inclination are similar in magnitude to the optimization without inclination. We are currently working on orbit configurations for a period of 20 years and with inclinations between $1°$ to $3°$

Table 2. Initial choice and adjusted choice of position x, y, z in unit of AU and velocity u, v, w in unit of AU/day for three S/C with 1° inclination

| S/C | Barycentric Rotating Frame | Barycentric Inertial Frame | Barycentric Rotating Frame (Adjusted) | Barycentric Inertial Frame (Adjusted) |
|---|---|---|---|---|
| 1 | -1.000001251450828 | -0.004802825764545 | -1.000001251450828 | -0.004802825764545 |
|   | 0.000000000000000 | 0.994627263012024 | 0.000000000000000 | 0.994627263012024 |
|   | 0.000000000000000 | 0.000158713293455 | 0.000000000000000 | 0.000158713293455 |
|   | 0.000000000000000 | -0.017192168339124 | 0.000000500000000 | -0.017192368180256 |
|   | 0.000002619968994 | -0.000007320653113 | 0.000002419968994 | -0.000007820716631 |
|   | -0.000300218712967 | -0.000300349354243 | -0.000300218712967 | -0.000300349354243 |
| 2 | 0.499996996517240 | 0.861567207169407 | 0.499996996517240 | 0.861567207169407 |
|   | 0.865893503923442 | -0.505095801449583 | 0.865893503923442 | -0.505095801449583 |
|   | 0.015114227203434 | 0.015272940496889 | 0.015114227203434 | 0.015272940496889 |
|   | -0.000002268956867 | 0.008602312820055 | -0.000002268956867 | 0.008602312820055 |
|   | -0.000001309974989 | 0.014898345843193 | -0.000001309974989 | 0.014898345843193 |
|   | 0.000150108266928 | 0.000149977625652 | 0.000150108266928 | 0.000149977625652 |
| 3 | 0.499996996517240 | -0.870219713271459 | 0.499996996517240 | -0.870219713271459 |
|   | -0.865893503923442 | -0.505646017047958 | -0.865893503923442 | -0.505646017047958 |
|   | -0.015114227203434 | -0.014955513909979 | -0.015114227203434 | -0.014955513909979 |
|   | 0.000002268956867 | 0.008611779134993 | 0.000002268956867 | 0.008611779134993 |
|   | -0.000001309974989 | -0.014896593527345 | -0.000001309974989 | -0.014896593527345 |
|   | 0.000150108266928 | 0.000149977625652 | 0.000150108266928 | 0.000149977625652 |

**4. Discussion and Conclusion**

In Section 2, the transfer orbits of ASTROD-GW formation have been designed under assumption of the spacecraft attracted by the Sun and the Earth only, and the delta-Vs are calculated through the finite difference method. Further studies will consider all the planets and Moon of the solar system. We also need take into accounts the inclination and the initial velocity adjustment of the mission orbits. The control forces shall also be taken in accounts, since the thrust performance of the propulsion module is limited.

In Section 3, the mission orbits with inclination and velocity adjustment have been simulated for the arm lengths evolutions to meet the measurement requirements.

In orbit design and orbit simulation for ASTROD I, we will deliver the S/C directly to the solar orbit and separate the S/C with the last stage of launcher in LEO (Low Earth Orbit) and use Venus gravity assistance.[9,33-35] We noted that for the Venus swing-by, to obtain the gravity assist to reach the other side of the Sun earlier, there is a launch window about every 584 days (synodic period of Venus).

If we use Venus flyby and Venus gravity assistance, after first spacecraft-Venus encounter the period of the spacecraft can be adjusted to Venus period, i.e. 224 days. At the first encounter, the spacecraft is ahead of the Earth by about 30 degrees in ecliptic longitude. At the second encounter after one round around the Sun, the spacecraft is ahead of the Earth by an additional 140 degrees. This time we use the thruster to adjust the second encounter with a small consumption of fuels so that the spacecraft will fly to an aphelion at about 1 AU at a position further ahead of the earth by 30 degrees. Altogether the spacecraft would be ahead of the earth by about 200 degrees. However, from our experience of orbital optimization of ASTROD I, longitude up to 30 degrees is usually adjustable in one encounter with a small consumption of fuels. Hence this option is good. Most fuels are for the last adjustment to stop in the final destination, in this case the L3 point of Sun and Earth. The propellant mass ratio should be lower and be in the range of 0.5-0.55. The total time to reach the destination would be around 1.3-1.5 years. For S/C 3, we could use a nearly Hohmann orbit with one revolution in 1.167 yr to transfer to near L5 point. The estimated propellant mass ratio needed would be around 0.365. We listed these estimates together with the ones for S/C 2 in Table 3. The transfer to near L4 destination could also be shortened to 0.833 yr with 1 revolution, if needed, and propellant mass ratio similar to the L3 transfer.

Table 3. Estimated Delta-V and Propellant Mass Ratio for Solar transfer of S/C

| S/C (Destination) | Transfer Orbit | Transfer Time | Solar Transfer Delta-V from beginning of nearly Hohmann orbit to destination | Solar Transfer Propellant Mass Ratio (Isp=320 s) |
|---|---|---|---|---|
| 1 (near L3) | Venus flyby transfer | 1.3-1.5 yr | 2.2-2.5 km/s | 0.50-0.55 |
| 2 (near L4) | Inner Hohmann, 2 Revolutions | 1.833 yr | 0.903 km/s | 0.250 |
| 3 (near L5) | Outer Hohmann, 1 Revolution | 1.167 yr | 1.422 km/s | 0.365 |

Table 3 gives an estimate of Delta-Vs and transfer times needed for deployment of the formation. The corresponding total mass for spacecraft 1, 2 and 3 are 1111-1266 kg, 690 kg, and 835 kg for a dry spacecraft (not including propeller and propellant) mass of 500 kg. Further studies on the optimizations of deployment from separation of launcher(s) in LEO(s) and on the orbit configurations for a period of 20 years and with inclinations between 1° to 3° are ongoing.

**Acknowledgements**

This work is supported in part by the National Science Council (under Grant No. NSC101-2112-M-007-007) and the National Center for Theoretical Sciences (NCTS).

**References**

1. Leitner, J., 2004. Formation flying—the future of remote sensing from space. In: Proceedings of the 18th International Symposium on Space Flight Dynamics. (ESA SP-548) Munich, Germany, 11–15 October.
2. GRACE, http://www.csr.utexas.edu/grace/.
3. GOCE, http://www.esa.int/SPECIALS/GOCE/index.html.
4. TechSat-21, http://space.skyrocket.de/doc_sdat/techsat-21.htm.
5. GPS, http://www.gps.gov/.
6. Iridium, http://www.iridium.com/default.aspx.
7. DMC, http://www.dmcii.com/about_us_constellation.htm.
8. FORMOSAT-3, http://www.nspo.org.tw/2008e/projects/project3/intro.htm.
9. H. Selig, C. Lämmerzahl and W.-T. Ni, Astrodynamical Space Test of Relativity using Optical Devices I (ASTROD I) - Mission Overview, *Int. J. Mod. Phys. D* **22** (2013) 1341003; and references therein.
10. LISA Study Team 2000 LISA (Laser Interferometer Space Antenna): A Cornerstone Mission for the Observation of Gravitational Waves, ESA System and Technology Study Report ESA-SCI(2000)11
11. O. Jennrich et al., "NGO -- Revealing a Hidden Universe: Opening a New Chapter of Discovery", Assessment Study Report, ESA/SRE[2011]19, ESA, December 2011.
12. S. Kawamura, M. Ando, N. Seto *et al*., *Class. Quantum Grav.* **28** (2011) 094011.
13. M. Ando and the DECIGO Working Group, DECIGO Pathfinder Mission, *Int. J. Mod. Phys. D* **22** (2013) 1341002; and references therein.
14. Crowder J and Cornish N J 2005 Phys. Rev. D 72 083005; and references therein.
15. W.-T. Ni, ASTROD-GW: Overview and Progress, *Int. J. Mod. Phys. D* **22** (2013) 1341004; and references therein.

16. W.-T. Ni, Dark energy, co-evolution of massive black holes with galaxies, and ASTROD-GW, *J. Adv. Space Res.* (2012), http://dx.doi.org/10.1016/j.asr.2012.09.019.
17. W.-T. Ni, Super-ASTROD: Probing Primordial Gravitational Waves and Mapping the Outer Solar System, *Class. Quantum Grav.* **26** (2009) 075021.
18. P. W. McNamara (on behalf of the LPF Team), The LISA Pathfinder Mission, *Int. J. Mod. Phys. D* **22** (2013) 1341001.
19. M. Ando, S. Kawamura, N.S Sato *et al*., DECIGO Pathfinder, *Class. Quantum Grav.*. **26** (2009) 094019.
20. The LIGO Scientific Collaboration, http://www.ligo.caltech.edu/
21. See also, C. S. Unnikrishnan, IndIGO and LIGO-India: Scope and Plans for Gravitational Wave Research and Precision Metrology in India, *Int. J. Mod. Phys. D* **22** (2013) 1341010 (2013).
22. W.-T. Ni, ASTROD and ASTROD I -- Overview and Progress, *Int. J. Mod. Phys. D* **17** (2008) 921-940; and references therein.
23. C. Grimani, Implications of Galactic and Solar Particle Measurements on Board Interferometers for Gravitational Wave Detection in Space, *Int. J. Mod. Phys. D* **22** (2013) 1341006.
24. R. Manchester, Pulsar Surveys and Timing, *Int. J. Mod. Phys. D* **22** (2013) 1341007; and references therein.
25. B. C. Joshi, Pulsar Timing Arrays, *Int. J. Mod. Phys. D* **22** (2013) 1341008; and references therein.
26. PPTA, http://www.atnf.csiro.au/research/pulsar/ppta/.
27. EPTA, http://www.epta.eu.org/.
28. NANOGRAV, http://nanograv.org/.
29. For a classification of GWs according to frequency, please see http://astrod.wikispaces.com/file/view/GW-classification.pdf
30. W.-T. Ni, Gravitational Waves, Dark energy and Inflation, *Mod. Phys. Lett. A* **25**, 922-935, 2010, and references therein; arXiv:1003.3899.
31. L. Collaztz, *The Numerical Treatment of Differential Equations*, Springer-Verlag, 1966.
32. An-Ming Wu, Xiaohui Xu and Wei-Tou Ni, Orbit Design and Analysis for ASTROD Mission Concept, *Int. J. Mod. Phys. D* **9** (2000) 201.
33. Claus Braxmaier, Hansjörg Dittus, Bernard Foulon, Ertan Göklü, Catia Grimani, Jian Guo, Sven Herrmann, Claus Lämmerzahl, Wei-Tou Ni, Achim Peters, Benny Rievers, Étienne Samain, Hanns Selig, Diana Shaul, Drazen Svehla, Pierre Touboul, Gang Wang, An-Ming Wu, and Alexander F. Zakharov, Astrodynamical Space Test of Relativity using Optical Devices I (ASTROD I) - A class-M fundamental physics mission proposal for Cosmic Vision 2015-2025: 2010 Update, *Exp. Astron.*, **34** (2012) 181[arXiv:1104.0060]; and references.
34. T. Appouchaux *et al.*, Astrodynamical Space Test of Relativity Using Optical Devices I (ASTROD I) - A class-M fundamental physics mission proposal for Cosmic Vision 2015-2025, Exp. Astron. 23 (2009) 491.
35. Ni, W.-T. *et al.*, ASTROD I: mission concept and Venus flybys. In: Proceedings of the 5th IAA International Conference on Low-Cost Planetary Missions, ESTEC, Noordwijk, The Netherlands, 24–26 September 2003, ESA SP-542, November 2003, pp.79–86; also: Acta Astronautica 59 (2006) 598.
36. Standish (1998). "JPL Planetary and Lunar Ephemerides, DE405/LE405". JPL Interoffice Memorandum 312.F-98-048.
37. J. R. Men, W.-T. Ni and G. Wang, Design of ASTROD-GW Orbit (in Chinese), *Acta Astron. Sin.* **51** (2009) 198.
38. J. R. Men, W.-T. Ni and G. Wang, Design of ASTROD-GW Orbit, *Chin. Astron. Astrophys.* **34** (2010) 434.
39. G. Wang and W.-T. Ni, ASTROD-GW time delay interferometry (in Chinese), *Acta Astron. Sin.* **52** (2011) 427.
40. G. Wang, and W.-T. Ni, ASTROD-GW time delay interferometry, *Chinese Astronomy and Astrophysics*, **36** ( 2012) 211-228.

41. G. Wang and W.-T. Ni, Orbit optimization for ASTROD-GW and its time delay interferometry with two arms using CGC ephemeris, *Chinese Physics B*, in press; arXiv:1205.5175.
42. V. A. Brumberg, *Essential Relativistic Celestial Mechanics* (Bristol: Adam Hilger, 1991), pp.178-177.